\documentclass[12pt]{article}

\usepackage{amsmath}
\usepackage{amsfonts}
\usepackage{eufrak}
\usepackage{amssymb,bbold}
\usepackage{epsfig}
\usepackage{bm}

\def\be{\begin{equation}}
\def\ee{\end{equation}}

\def\a{\alpha}
\def\b{\beta}

\def\v{\nu}

\def\m{\mu}
\def\s{\sigma}

\def\e{\epsilon}
\def\G{\Gamma}
\def\md{\mathcal{D}}

\def\d{\delta}

\def\6{\partial}
\def\mbb{\mathbb{R}}

\newcommand{\ip}{\raise1pt\hbox{\large$\lrcorner$}\,}
\newcommand{\bea}{\begin{eqnarray}}
\newcommand{\eea}{\end{eqnarray}}
\newcommand{\nn}{\nonumber \\}

\begin{document}
\title{Null structure groups in eleven dimensions}

\author{Marco Cariglia \thanks{M.Cariglia@damtp.cam.ac.uk} and
  Ois\'{\i}n A. P. Mac
  Conamhna\thanks{O.A.P.MacConamhna@damtp.cam.ac.uk} \\ DAMTP \\ Centre
  for Mathematical Sciences \\ University of Cambridge \\ Wilberforce
  Road, Cambridge CB3 0WA, UK.}

\maketitle 

\abstract{We classify all the structure groups which arise as
  subgroups of the isotropy group,
  $(Spin(7)\ltimes\mathbb{R}^8)\times\mathbb{R}$, of a single null
  Killing spinor in eleven dimensions. We construct the spaces of spinors
  fixed by these groups. We determine the conditions under which
  structure subgroups of the maximal null strucuture group
  $(Spin(7)\ltimes\mathbb{R}^8)\times\mathbb{R}$ may also be embedded
  in $SU(5)$, and hence the conditions under which a supersymmetric
  spacetime admits only null, or both timelike and null, Killing
  spinors. We discuss how this purely algebraic material will
  facilitate the direct analysis of the Killing spinor equation of
  eleven dimensional supergravity, and the classification of
  supersymmetric spacetimes therein.} 

\section{Introduction}
Recent advances in the understanding of supergravity theories and
their supersymmetric solutions 
have reached the stage where geometries solving the Killing spinor
equation can be systematically classified and constructed. The key
concept which renders this possible is that of a G-structure. In this work we 
undertake the task of finding all the possible structure
groups defined by a set of Killing spinors, at least one of which is null, which can arise in eleven dimensional supergravity, and explicitly
constructing the spaces
of spinors that these groups preserve. Our motivation is ultimately to construct a road
map to eleven dimensional null supersymmetry: we want a recipe to
associate to any 
class of solution admitting a null Killing spinor, be it $M2$--branes with transverse $Spin(7)$
space, $G_2$ compactifications to four dimensions, more complex systems of
intersecting membranes and so on, a well defined space of candidate
Killing spinors and a set of rules to calculate all the restrictions
given by supersymmetry on the geometry and the fluxes. The method that
we will employ can be adapted without conceptual modifications
to lower dimensional supergravities.
 
There are a number of reasons why we believe this undertaking to be of
interest, of
which we mention three in particular. First, a
systematic understanding of the structure of supersymmetric solutions
gives powerful tools for constructing new and interesting ones. An
eloquent example is the discovery of black ring solutions in five dimensional
supergravity \cite{blackring}, which has stimulated discussion on the
non-uniqueness of supersymmetric asymptotically flat black-hole
solutions and on M--Theory calculations of their entropy. Second, there
is a natural desire to classify supersymmetric solutions, for example
to understand which flux-compactifications to lower
dimensions are admissible in string and M-theory
\cite{J1}-\cite{minasian}, and more generally, to provide as detailed
a map as possible of the entire landscape of supersymmetric spacetimes
in M-theory. Third, there is the potential for advances in Riemannian
geometry; for example, the construction of a new infinite family of
Einstein-Sasaki manifolds has already emerged from the G-structure
programme \cite{einstein1}, \cite{einstein2}.
 
Since we are going to employ a refinement of the standard G-structure
approach, we will briefly review the development of the
method itself. A direct precursor of the G--structure analysis is
\cite{Tod}. The utility of G-structures in the context of classifying
supersymmetric spacetimes was first advocated and demonstrated in
\cite{gmpw}. In \cite{gaunt} it was shown how
powerful this technique can be when applied to simple lower
dimensional supergravities. By now there exists an extensive
literature on applications of G--structure techniques to supergravity
\cite{saffin}-\cite{us}. The first papers 
to study eleven dimensional supergravity with G--structure methods are \cite{gaunt1},
\cite{gaunt3}. There necessary and sufficient
conditions are given on the metric and four--form for the existence of either a
single timelike or a single 
null Killing spinor. The existence of such spinors is
equivalent to the existence of an $SU(5)$ or
$(Spin(7)\ltimes\mbb^8)\times\mbb$ structure, respectively, and the
constraints for supersymmetry are encoded in the intrinsic
torsion. The drawback of the original approach was that it could be
only used for the classification of spacetimes with minimal
supersymmetry. However, following the suggestion of \cite{gaunt1}, a
systematic and universally applicable refinement of the original
method, capable of classifying all solutions admitting an arbitrary
number of Killing spinors, was given in \cite{Ois}, and
afterwards \cite{pap}.
 
The contributions of \cite{gaunt1,gaunt3} represent a major
theoretical advance because for the first time the study of
supersymmetric solutions in eleven dimensions was based on a systematic
understanding of all the geometries that support a solution of the
Killing equation. The only alternative known before (with the
exception of spacetimes with maximal supersymmetry) was that of
giving an ansatz for the solutions and then verifying supersymmetry a
posteriori. Mathematically a G-structure on a spacetime $M$ is a
principal sub-bundle of the frame 
bundle of $M$. A spacetime has holonomy G if it admits a torsion-free
G-structure. In more concrete terms, the deviation from holonomy of a
supersymmetric spacetime induced by the presence of fluxes is encoded
in the intrinsic torsion of the G-structure (a general discussion of
G-structures is given, for example, in \cite{joyce}). It is this that 
makes a G-structure such a useful concept in the classification of
supersymmetric spacetimes. 

However, the nature of eleven dimensional supergravity is such
that the space of backgrounds on which the Killing equation can be
solved is vast. Trying to start
from any such general background and to impose the
equations of motion of the theory on it seems at present a prohibitive
task. A natural and physically interesting specialization to make is
to the study of geometries with enhanced supersymmetry, where there are two or
more linearly independent solutions of the Killing
equation. Classifying all such geometries is what we mean by a refined
G--structure classification. Attempting to do this in the original
G-structure formalism, which worked at the level of the bilinears
constructed from the Killing spinors, is technically very complicated.
In \cite{Ois} a new approach for
solving this problem was proposed, and illustrated in seven
dimensions, in which all
calculations are performed on the spinors themselves rather than their
bilinears (the formalism includes and generalises the algebraic
Killing spinor techniques of \cite{warn1}-\cite{warn5}). Applying this 
method to any other supergravity is 
straightforward. 
 
The essential point which makes calculations tractable is the
simplified description of the spinors.  It is
shown in \cite{Ois} that by
acting on a fiducial spinor $\e$ with matrices $Q$ in an appropriate subset
of the Clifford algebra, one may span the space of spinors, in any
supergravity. One then decomposes the space of spinors into modules of
the maximal compact factor of the isotropy group of the fiducial spinor
$\e$, and likewise the spinorial basis into bases for the modules. In
eleven dimensions, the appropriate groups are $SU(5)$ or
Spin(7). The most efficient analysis of the Killing
spinor equation then proceeds as follows, and may be broken into three phases.

First, one assumes that the fiducial spinor $\e$ is Killing, and
calculates the constraints implied by its existence. These conditions
may be easily 
computed, using the standard G-structure formalism (as was done for
eleven dimensional supergravity in \cite{gaunt1, gaunt3}), or, for maximal
computational efficiency, the streamlined versions thereof given in
\cite{minasian}, \cite{Ois}. Then since $\e$ is Killing, its isotropy
group is promoted to the status of the preferred local structure group
of the spacetime, and one has already constructed a spinorial basis
decomposed into modules of the maximal compact factor of this structure group.

Second, one determines all the possible preferred local structure
groups defined by arbitrary sets of additional Killing spinors, and
the spaces of spinors they fix. This gives a classification of the
possible preferred local G-structures admitted by supersymmetric
spacetimes in the supergravity, together with the number of
supersymmetries consistent with each. The main objective of this paper
is to perform this classification for sets of Killing
spinors in eleven dimensions which contain at least one null Killing
spinor but which are otherwise arbitrary.

Third, one computes the additional conditions on the intrinsic torsion
and fluxes implied by the existence of an arbitrary additional Killing
spinor $\eta$. Since $\eta$ is
of the form $\eta=Q\e$, $\eta$ is Killing if and only if
\be\label{big}
[\md_{\m},Q]\e=0, 
\end{equation}
where $\md_{\m}$ is the supercovariant derivative. By imposing the
defining projections satisfied by the fiducial Killing spinor $\e$,
the spinor $[\md_{\m},Q]\e$ may be written as a manifest sum of basis 
spinors, and by linear independence, the coefficient of each must
vanish separately. The most convenient and concise way of expressing
these conditions is to further exploit the existence of the
G-structure defined by $\e$, and to decompose the spin connection and
fluxes into modules of the maximal compact factor of the structure group.  
In this way the Killing spinor equation for $\eta$ is converted into a
set of purely bosonic equations for tensors of the maximal compact
factor of the structure group defined by $\e$, which give additional
constaints on the intrinsic torsion and fluxes. Once the general conditions
for the existence of an arbitrary additional Killing spinor have been
computed, the conditions implied by the existence of any particular
set may be analysed in detail. In a forthcoming series of papers
\cite{spin7,y,z}, we will compute the conditions implied by (\ref{big})
for general $Q$ and null $\e$ in eleven dimensions.

In \cite{pap}, some of the structure groups arising as subgroups
of $SU(5)$ in eleven dimensions were classified. All such possible
structure groups 
which can fix two Killing spinors, and some special cases of
structure groups fixing $N>2$, were given. The constraints on the
bosonic fields 
associated with all $N=2$ $SU(5)$ structures and some special cases of
$SU(4)$ structures were computed. The description of spinors in
\cite{pap} exploits the 
isomorphism between the space of Dirac spinors in eleven dimensional
spacetime, and the space of $(0,p)$ forms, $p=0,...,5$, in ten
Riemannian dimensions. Though this language was not used explicitly in
\cite{Ois}, the description of the spinors is manifestly equivalent;
the authors of \cite{pap} construct the basis in spinor space by
acting on a fiducial complex (timelike) spinor $\rho$ with
\be
R_{a_1...a_n}\G^{a_1...a_n},
\end{equation}
where the $R_{a_1...a_n}$ given a basis for $(0,p)$ forms in ten
Riemannian dimensions. By supressing $\rho$ and the Gamma-matrices,
the spinors in \cite{pap} are then treated as forms throughout. As was
done in \cite{Ois}, this spinorial basis is then used to convert the Killing
spinor equation into a set of purely bosonic equations.

As stated above, the objective of this work is to give the
classification of all possible structure groups 
of supersymmetric eleven dimensional spacetimes which 
may be embedded in $(Spin(7)\ltimes\mbb^8)\times\mbb$, and to
construct the spaces of spinors fixed by these groups. Since so many
distinct cases are involved, the refined G-structure classification of supersymmetric
spacetimes in eleven dimensions is a problem of great
proportions. Here we will lay the algebraic groundwork for this
problem, for all spacetimes with structure groups embedding in
$(Spin(7)\ltimes\mbb^8)\times\mbb$. In section two
we will construct a basis of spinors by acting on a fiducial null
spinor with a subset of the Clifford algebra, and decompose the basis into
modules of Spin(7). In section three we will
use this basis to determine the possible structure groups, and the
spaces of spinors they fix. In section four we discuss the
structure groups which may be embedded both in
$(Spin(7)\ltimes\mbb^8)\times\mbb$ 
and $SU(5)$ (so that the spacetimes with these structures admit both
timelike and null Killing spinors), and determine the conditions on the
spinors for this to occur. In particular, we will find some possible structure
groups which may be embedded in $SU(5)$ which are not covered by
\cite{pap}. Section five contains our conclusions. We also provide a
sample of the sort of results that can be expected using our method
\cite{spin7}: the general local bosonic solution of the
Killing spinor equation in eleven dimensions, admitting a Spin(7)
structure. 

\section{The spinor basis}
As discussed in the introduction, the most efficient way of analysing
the Killing spinor equation with more than one supersymmetry is to act
on a  Killing spinor with a subset of the Clifford algebra, so
  that we may use (\ref{big}). Since the constraints associated with
  the existence of a single null Killing spinor were computed in
  \cite{gaunt3}, we will adopt all their conventions so that their
  results may be readily incorporated into the refined
  classification. Before constructing the spinor basis, and to fix 
  our notation, let us first summarise some useful material. In the
  spacetime basis
\be
ds^2=2e^+e^-+e^ie^i+e^9e^9,
\end{equation}
where $i=1,...,8$, and where the eight dimensional manifold spanned by the
$e^i$ will be referred to as the base, a single null spinor $\e$ may be fixed by the projections
\bea
\G_{1234}\e=\G_{3456}\e=\G_{5678}\e=\G_{1357}\e&=&-\e,\nn\G^+\e&=&0.
\eea
The only non-zero bilinears which may be constructed from this spinor
are the one form, two form and five form which are 
\bea
K&=&e^+,\nn \Omega&=&e^+\wedge e^9,\nn\label{bilin}\Sigma&=&e^+\wedge\phi,
\eea
where $\phi$ is the Spin(7) four form whose only non-zero components
are given by
\bea
-\phi&=&e^{1234}+e^{1256}+e^{1278}+e^{3456}+e^{3478}+e^{5678}+e^{1357}\nn&-&e^{1368}-e^{1458}-e^{1467}-e^{2358}-e^{2367}-e^{2457}+e^{2468}. 
\eea
This Spin(7) subgroup of Spin(8) is that which splits the positive
chirality irrep $\mathbf{8}_{+}$ of Spin(8) into
$\mathbf{8}_+=\mathbf{1}+\mathbf{7}$ while leaving the negative chirality
and vector irreps $\mathbf{8}_-$, $\mathbf{8}_v$ irreducible. Under
Spin(7), the adjoint of Spin(8) splits as
$\mathbf{28}=\mathbf{21}+\mathbf{7}$. For two forms, the projections may be
written as
\bea
(P^{\mathbf{21}}A)_{ij}&=&\frac{3}{4}(A_{ij}+\frac{1}{6}\phi_{ijkl}A^{kl}),\nn(P^{\mathbf{7}}A)_{ij}&=&\frac{1}{4}(A_{ij}-\frac{1}{2}\phi_{ijkl}A^{kl}).
\eea
We may choose a basis for the $\mathbf{7}$ to be
\bea
J^1=e^{18}+e^{27}-e^{36}-e^{45},& &J^2=e^{28}-e^{17}-e^{35}+e^{46},\nn
J^{3}=e^{38}+e^{47}+e^{16}+e^{25},&&J^4=e^{48}-e^{37}+e^{15}-e^{26}\nn
J^{5}=e^{58}+e^{67}-e^{14}-e^{23},&&J^6=e^{68}-e^{57}-e^{13}+e^{24},\nn
J^7=e^{78}+e^{56}&+&e^{34}+e^{12}.
\eea
They obey
\bea
J^A_{ik}J^{Bk}_{\;\;\;\;\;j}&=&-\d^{AB}\d_{ij}+K^{AB}_{ij},\nn
J^{Aij}K^{BC}_{ij}&=&0,\nn\label{alg}
K^{ABij}K^{CD}_{ij}&=&8(\d^{AC}\d^{BD}-\d^{AD}\d^{BC}),
\eea
where $A,B=1,...,7$ and the $K^{AB}_{ij}$ are antisymmetric on $A,B$
and furnish a basis for the $\mathbf{21}$. Note that
\bea\label{ann}
K^{AB}_{ij}\G^{ij}\e&=&0,\\\label{annh}J^{A}_{ij}\G^{ij}\e&=&8\G^{A8}\e.
\eea 
The most general element of the Lie algebra of Spin(1,10) which
annihilates $\e$ is thus obviously
\be\label{lie}
f^{AB}K^{AB}_{ij}\G^{ij}+\theta_i\G^{+i}+q\G^{+9}
\end{equation}
which of course generates $(Spin(7)\ltimes\mbb^8)\times\mbb$.
Now we turn to the construction of the spinor basis. We define a basis of spinors by the projections (with no sum on $i$)
\bea\label{proj}
\G_{1234}\e_{(i)}&=&-\a^1_{(i)}\e_{(i)},\nn
\G_{3456}\e_{(i)}&=&-\a^2_{(i)}\e_{(i)},\nn
\G_{5678}\e_{(i)}&=&-\a^3_{(i)}\e_{(i)},\nn
\G_{1357}\e_{(i)}&=&-\a^4_{(i)}\e_{(i)},\nn \G^{\a^5_{(i)}}\e_{(i)}&=&0,
\eea
for the thirty-two possible combinations of
$\a^{1,...,5}_{(i)}=\pm1$. We want to construct this basis by acting
on our fiducial spinor $\e\equiv\e_{(1)}$ (defined by
$\a^{1,...,5}_{(1)}=+$) with a subset of the Clifford algebra. First
consider the sixteen spinors with $\a^5_{(i)}=+$. We may split these
spinors into two groups of eight, according to whether their chirality
on the eight dimensional base is positive or negative (this chirality
is given by $\a^1_{(i)}\a^3_{(i)}$). We may easily solve the
projections (\ref{proj}) for the eight basis spinors with
$\a^5_{(i)}=+$ and chirality $\a^1_{(i)}\a^3_{(i)}=+$ to find that
they are given by
\be
\e,\;\;J^A_{ij}\G^{ij}\e,
\end{equation}
where $A=1,...,7$ and the splitting of the $\mathbf{8}_+$ of Spin(8) into the
$\mathbf{1}+\mathbf{7}$ of Spin(7) is manifest. Next the eight basis
spinors with $\a^5_{(i)}=+$, $\a^1_{(i)}\a^3_{(i)}=-$ are readily
found to be
\be
\G^i\e.
\end{equation}
The eight basis spinors with $\a^5_{(i)}=-$, $\a^1_{(i)}\a^3_{(i)}=+$
are given by
\be
\G^-\e,\;\;J^A_{ij}\G^{-ij}\e,
\end{equation}
and the eight basis spinors with $\a^5_{(i)}=-$,
$\a^1_{(i)}\a^3_{(i)}=-$ are
\be
\G^{-i}\e.
\end{equation}
Thus an arbitrary Majorana spinor $\eta$ in eleven dimensions may be written as
\be
\eta=(f+\frac{1}{8}f^AJ^A_{ij}\G^{ij}+u_i\G^i+g\G^-+\frac{1}{8}g^AJ^A_{ij}\G^{-ij}+v_i\G^{-i})\e,
\end{equation}
for thirty-two real functions $f$, $f^A$, $u_i$, $g$, $g^A$, $v_i$,
and the factors of $1/8$ are introduced for future convenience. This equation is manifestly
equivalent to the statement (in the language of \cite{pap}) that
the space of Majorana spinors in eleven dimensions is isomorphic to
the direct sum of the spaces of Spin(7) forms, defined on the eight
dimensional base,
\be
\Lambda^0_1\oplus\Lambda^0_1\oplus\Lambda^1_8\oplus\Lambda^1_8\oplus\Lambda^2_7\oplus\Lambda^2_7.
\end{equation}

\section{The structure groups}
In this section we will classify the possible structure groups arising
as subgroups of $(Spin(7)\ltimes\mbb^8)\times\mbb$, and the spaces
of spinors they fix. This will give a complete classification of all
G-structures which can arise for eleven dimensional spacetimes
admitting at least one null Killing spinor, and will organise the
enormously complicated problem of analysing the Killing spinor
equation into many, more manageable, sub-problems. We will henceforth
assume that the fiducial spinor $\e$ is Killing, and we will determine
the reduction of the structure group implied by the existence of all
other sets of Killing spinors. An important point we will
exploit throughout is our freedom to perform
$(Spin(7)\ltimes\mbb^8)\times\mbb$ transformations preserving
the metric and leaving the fiducial spinor $\e$ (and hence the
bilinears (\ref{bilin})) invariant. This freedom will allow us to take
some additional Killing spinors to be of a simple, ``canonical'' form;
this point was also exploited in \cite{pap}. Let us also comment on
our notation: throughout this section, the index $A$ takes the values
1,...,7. $B$ takes values 1,...,6, $C\in\{1,...,5\}$,
$D\in\{1,...,4\}$, $E\in\{1,2,3\}$, $F\in\{1,2\}$, $G\in\{6,7\}$,
  $H\in\{5,6,7\}$, $I\in\{4,...,7\}$. 

\subsection{Additional Killing spinors of the form
  $(f+\frac{1}{8}f^AJ^A_{ij}\G^{ij})\e$}
In this subsection we will explore the reduction of the structure
group implied by the existence of progressively more Killing spinors
of the form $(f+\frac{1}{8}f^AJ^A_{ij}\G^{ij})\e$.  The structure groups which can arise by
demanding Killing spinors of this form, together with the number of
Killing spinors of this form they can fix, are summarised as follows:
\begin{center}
\begin{tabular}{|c|c|}\hline$G$&$N$\\\hline\hline
  $(SU(4)\ltimes\mbb^8)\times\mbb$&2\\\hline $(Sp(2)\ltimes\mbb^8)\times\mbb$&3\\\hline
  $(SU(2)\times SU(2))\ltimes\mbb^8)\times\mbb$&4\\\hline 
  $(SU(2)\ltimes\mbb^8)\times\mbb$&5\\\hline $(U(1)\ltimes\mbb^8)\times\mbb$&6\\\hline
  $\mbox{Chiral}\;\;\mbb^9$&7,8\\\hline
\end{tabular}
\end{center}
By something of an abuse of language, we take a ``chiral $\mbb^9$
structure'' to refer to a G-structure defined by a set of Killing
spinors with common isotropy group $\mbb^9$ which share the same
chirality on the eight dimensional base. Let us now consider each case
in more detail.

\paragraph{$(SU(4)\ltimes\mbb^8)\times\mbb$ structure, N=2.}
Suppose we  demand the existence of a second linearly independent Killing
spinor of the form
\be
\eta_{(1)}=(f_{(1)}+\frac{1}{8}f_{(1)}^AJ^A_{ij}\G^{ij})\e.
\end{equation}
 We will now exploit the fact that
Spin(7) acts transitively on the unit sphere in the $\mathbf{7}$. We
may perform a Spin(7) transformation, preserving the metric, $\e$ and
the bilinears, to
write $\eta_{(1)}$ as
\be\label{su4}
\eta_{(1)}=(f_{(1)}+\frac{1}{8}g_{(1)}J^7_{ij}\G^{ij})\e,
\end{equation}
where $(g_{(1)})^2=f^A_{(1)}f^A_{(1)}$. Since
\be
\frac{Spin(d+1)}{Spin(d)}\cong S^{d},
\end{equation}
the stabiliser of $\eta_{(1)}$ in Spin(7) is Spin(6)$\cong
SU(4)$, and the existence of this Killing spinor implies that the structure
group is reduced to $(SU(4)\ltimes\mbb^8)\times\mbb$. This may be
verified directly by applying the general Lie algebra element
(\ref{lie}) to $\eta_{(1)}$. Using (\ref{alg}) and (\ref{ann}), it is
straightforward to verify that $\eta_{(1)}$ is annihilated if and only
if $f^{7A}=0$ in (\ref{lie}). The common isotropy algebra of $\e$ and
$\eta_{(1)}$ is thus that of $(SU(4)\ltimes\mbb^8)\times\mbb$, and
this structure group fixes precisely two spinors. $\eta_{(1)}$ is
linearly independent of $\e$ if $g_{(1)}\neq0$, but the most generic
additional spinor consistent with an $(SU(4)\ltimes\mbb^8)\times\mbb$
structure has $f_{(1)}\neq0$.

\paragraph{$(Sp(2)\ltimes\mbb^8)\times\mbb$ structure, N=3.}
Suppose we demand, in addition to $\e$ and $\eta_{(1)}$, the existence
of a third
linearly independent Killing spinor of the form
\be
\eta_{(2)}=\eta_{SU(4)\ltimes}+\frac{1}{8}f_{(2)}^BJ^B_{ij}\G^{ij}\e,
\end{equation}
where $\eta_{SU(4)\ltimes}$ is an arbitrary spinor fixed by the
$(SU(4)\ltimes\mbb^8)\times\mbb$ of the previous paragraph, and
$1/8f_{(2)}^BJ^B_{ij}\G^{ij}\e$, $B=1,...,6$ is an arbitrary non-zero spinor in the
$\mathbf{6}$ of Spin(6). Using the fact
that Spin(6) acts transitively on the unit sphere its $\mathbf{6}$ with stabiliser
Spin(5) $\cong Sp(2)$, we may write $\eta_{(2)}$ as
\be\label{sp2}
\eta_{(2)}=\eta_{SU(4)\ltimes}+\frac{1}{8}g_{(2)}J^6_{ij}\G^{ij}\e,
\end{equation}
and deduce that the common isotropy group of $\e$, $\eta_{(1)}$ and
$\eta_{(2)}$ is $(Sp(2)\ltimes\mbb^8)\times\mbb$. Together with the other two
Killing spinors, $\eta_{(2)}$ is annihilated if and only if in
addition $f^{6A}=0$
in (\ref{lie}). 

\paragraph{$((SU(2)\times SU(2))\ltimes\mbb^8)\times\mbb$ structure,
  N=4.}
Suppose we demand the existence of a fourth linearly independent
Killing spinor of the form
\be
\eta_{(3)}=\eta_{Sp(2)\ltimes}+\frac{1}{8}f_{(3)}^CJ^C_{ij}\G^{ij}\e,
\end{equation}
where $\eta_{Sp(2)\ltimes}$ is an arbitrary spinor fixed by the
$(Sp(2)\ltimes\mbb^8)\times\mbb$ of the previous paragraph, and
$1/8f_{(3)}^CJ^C_{ij}\G^{ij}\e$, $C=1,...,5$ is a non-zero spinor in the
$\mathbf{5}$ of Spin(5). Since Spin(5)
acts transitively on the unit sphere in its $\mathbf{5}$ with stabiliser
Spin(4) $\cong SU(2)\times SU(2)$, we may write $\eta_{(3)}$ as
\be\label{su2times}
\eta_{(3)}=\eta_{Sp(2)\ltimes}+\frac{1}{8}g_{(3)}J^5_{ij}\G^{ij}\e,
\end{equation}
and deduce that the existence of this Killing spinor in addition to
the previous three implies that the structure group is reduced to
$((SU(2)\times SU(2))\ltimes\mbb^8)\times\mbb$. Together with the other
three Killing spinors, $\eta_{(3)}$ is annihilated if and only if
in addition $f^{5A}=0$ in (\ref{lie}).

\paragraph{$(SU(2)\ltimes\mbb^8)\times\mbb$ structure, N=5.}  
Suppose we demand the existence of a fifth linearly independent
Killing spinor of the form
\be
\eta_{(4)}=\eta_{SU(2)\times SU(2)\ltimes}+\frac{1}{8}f_{(4)}^DJ^D_{ij}\G^{ij}\e,
\end{equation}
where $\eta_{SU(2)\times SU(2)\ltimes}$ is an arbitrary spinor fixed
by the $((SU(2)\times SU(2))\ltimes\mbb^8)\times\mbb$ of the previous
paragraph and $1/8f_{(4)}^DJ^D_{ij}\G^{ij}\e$, $D=1,...,4$ is a non-zero
spinor in the $\mathbf{4}$ of Spin(4). Since Spin(4)
acts transitively on the unit sphere in its $\mathbf{4}$ with stabiliser
Spin(3) $\cong SU(2)$, we may write $\eta_{(4)}$ as
\be\label{su2}
\eta_{(4)}=\eta_{SU(2)\times SU(2)\ltimes}+\frac{1}{8}g_{(4)}J^4_{ij}\G^{ij}\e,
\end{equation}
and deduce that the existence of this Killing spinor in addition to
the previous four implies that the structure group is reduced to
$(SU(2)\ltimes\mbb^8)\times\mbb$. Together with the other
four Killing spinors, $\eta_{(4)}$ is annihilated if and only if
in addition $f^{4A}=0$ in (\ref{lie}). Note that the $SU(2)$ factor
in this subgroup of $(Spin(7)\ltimes\mbb^8)\times\mbb$ acts
nontrivially on all eight dimensions of the base. It is certainly not
true that a holonomy group that is strictly of this form (and not some
subgroup) can arise, since an $SU(2)$ holonomy acting nontrivially on
eight dimensions is not on Berger's list. In the abscence of flux
(so the intrinsic torsion of the G-structure vanishes),
the existence of the Killing spinor $\eta_{(4)}$ in addition to the
other four would identically imply the existence of more Killing
spinors, which would reduce the holonomy group to $\mbb^9$, with
$N=16$. When the 
flux is non-zero, the intrinsic torsion of the G-structure is
non-zero, and we are unaware of any theorems forbidding an $SU(2)$
structure (as opposed to holonomy) in eight dimensions. However, what
we are doing in this paper is purely algebraic, and it may well be
that a direct analysis of the Killing spinor equation will reveal that
the existence of the Killing spinor $\eta_{(4)}$ in addition to the
other four will always identically imply the existence of more
Killing spinors, so
that the structure group would be further reduced, and
strictly $(SU(2)\ltimes\mbb^8)\times\mbb$ structures would not
arise.   

\paragraph{$(U(1)\ltimes\mbb^8)\times\mbb$ structure, N=6.}  
Suppose we demand the existence of a sixth linearly independent
Killing spinor of the form
\be
\eta_{(5)}=\eta_{SU(2)\ltimes}+\frac{1}{8}f_{(5)}^EJ^E_{ij}\G^{ij}\e,
\end{equation}
where $\eta_{SU(2)\ltimes}$ is an arbitrary spinor fixed by the
$(SU(2)\ltimes\mbb^8)\times\mbb$ of the previous paragraph, and
$\frac{1}{8}f_{(5)}^EJ^E_{ij}\G^{ij}\e$, $E=1,2,3$ is a non-zero spinor in the
$\mathbf{3}$ of Spin(3). Since Spin(3)
acts transitively on the unit sphere in its $\mathbf{3}$ with stabiliser
Spin(2) $\cong U(1)$, we may write $\eta_{(5)}$ as
\be\label{u1}
\eta_{(5)}=\eta_{SU(2)\ltimes}+\frac{1}{8}g_{(5)}J^3_{ij}\G^{ij}\e,
\end{equation}
and deduce that the existence of this Killing spinor in addition to
the previous five implies that the structure group is reduced to
$(U(1)\ltimes\mbb^8)\times\mbb$. Together with the other
five Killing spinors, $\eta_{(5)}$ is annihilated if and only if
in addition $f^{3A}=0$ in (\ref{lie}). The same remarks as those made
regarding the existence or otherwise of the 
$(SU(2)\ltimes\mbb^8)\times\mbb$ structure apply to this case.

\paragraph{Chiral $\mbb^9$ structure, N=7,8.}   
Suppose we demand the existence of a seventh linearly independent
Killing spinor of the form
\be
\eta_{(6)}=\eta_{U(1)\ltimes}+\frac{1}{8}f_{(6)}^FJ^F_{ij}\G^{ij}\e,
\end{equation}
where $\eta_{U(1)\ltimes}$ is an arbitrary spinor fixed by the
$(U(1)\ltimes\mbb^8)\times\mbb$ of the previous paragraph, and
$1/8f_{(6)}^FJ^F_{ij}\G^{ij}\e$, $F=1,2$ is a non-zero spinor in the
$\mathbf{2}$ of Spin(2). Since Spin(2)
acts transitively on the unit sphere in its $\mathbf{2}$ with stabiliser
the identity, we may write $\eta_{(6)}$ as
\be\label{r9}
\eta_{(6)}=\eta_{U(1)\ltimes}+\frac{1}{8}g_{(6)}J^{2}_{ij}\G^{ij}\e,
\end{equation}
and deduce that the existence of this Killing spinor in addition to
the previous six implies that the structure group is reduced to
$\mbb^9$. Together with the other
six Killing spinors, $\eta_{(6)}$ is annihilated if and only if
$f^{AB}=0$ in (\ref{lie}). We have used the last of the Spin(7)
freedom to fix $\eta_{(6)}$ in the form (\ref{r9}). We can have one
further linearly independent Killing spinor with a chiral $\mbb^9$
structure. This is
$\eta_{(7)}=(f_{(7)}+\frac{1}{8}f_{(7)}^AJ^A_{ij}\G^{ij})\e$, $f_{(7)}^1\neq0$. We have no freedom
left to fix any of the components of this spinor while preserving the
form of the first seven, so generically all eight functions in
$\eta_{(7)}$ are nonzero.   

\subsection{Additional Killing spinors of the form
  $(f+\frac{1}{8}f^AJ^A_{ij}\G^{ij}+u_i\G^i)\e$} 
In this subsection we will examine the reduction of the structure
group implied by incorporating progressively more Killing spinors of
the more generic form $(f+\frac{1}{8}f^AJ^A_{ij}\G^{ij}+u_i\G^i)\e$. The
structure groups which can arise (distinct from those of the previous
subsection, which are obtained  with $u_i=0$), and the number of
Killing spinors they can fix, are summarised as follows:
\begin{center}
\begin{tabular}{|c|c|}\hline$G$&$N$\\\hline\hline
  $(G_2\ltimes\mbb^7)\times\mbb^2$&2\\\hline 
  $(SU(3)\ltimes\mbb^6)\times\mbb^3$&2,3,4\\\hline $(SU(2)\ltimes\mbb^4)\times\mbb^5$&3,4,...,8\\\hline
  $\mbb^9$&3,4,...,16\\\hline
\end{tabular}
\end{center}
Let us discuss each case in more detail. We point out that we reserve
the notation $\s$ for Killing spinors which are completely generic
among spinors of the form
discussed in this 
subsection. $\s_{(i)}$ denotes the $i$th such Killing spinor.

\paragraph{$(SU(3)\ltimes\mbb^6)\times\mbb^3$ structure, N=2,3,4.}
Suppose we demand the existence of a second
independent Killing spinor, of the most generic form discussed in this
subsection, 
\be\label{g2g2}
\s_{(1)}=(f_{(1)}+\frac{1}{8}f^A_{(1)}J^A_{ij}\G^{ij}+u_{(1)i}\G^i)\e.
\end{equation}
where $f^A_{(1)},\;u_{(1)i}\neq0$. We may use the
fact that Spin(7) acts transitively on the unit sphere in its
$\mathbf{8}_-$, and that
\be
\frac{Spin(7)}{G_2}\cong S^7,
\end{equation}
to deduce that we may take 
\be
\s_{(1)}=(f_{(1)}+\frac{1}{8}f^A_{(1)}J^A_{ij}\G^{ij}+u_{(1)}\G^8)\e,
\end{equation}
where $(u_{(1)})^2=u_{(1)i}u_{(1)}^i$, and we still have the freedom to
perform $G_2$ transformations preserving
$u_{(1)i}=u_{(1)}\d_{i8}$. Under $G_2$, the $\mathbf{8}_-$ of Spin(7)
decomposes as $\mathbf{8}_-=\mathbf{1}+\mathbf{7}_{G_2}$. 
The $\mathbf{7}$ of Spin(7) is left irreducible under
$G_2$. We now use the fact that $G_2$ acts transitively on the unit
sphere in its $\mathbf{7}$, and that 
\be
\frac{G_2}{SU(3)}\cong S^6,
\end{equation}
to deduce that we may take
\be
\s_{(1)}=(f_{(1)}+\frac{1}{8}g_{(1)}J^7_{ij}\G^{ij}+u_{(1)}\G^8)\e,
\end{equation}
and the
existence of $\s_{(1)}$ implies that the structure group is reduced
to $(SU(3)\ltimes\mbb^6)\times\mbb^3$, where the $SU(3)$ acts
nontrivially in the 123456 directions. The $\mathbf{7}$ of $G_2$
decomposes under $SU(3)$ as $\mathbf{7}=\mathbf{1}+\mathbf{3}+\overline{\mathbf{3}}$. Using (\ref{annh}), we note
that $\s_{(1)}$ is annihilated by (\ref{lie}) if and only if
$f^{AB}K^{AB}_{i8}=f^{AB}K^{AB}_{i7}=0$. These are thirteen
independent conditions, agreeing with the fact that the Lie algebra of
$SU(3)$ is eight dimensional. Finally, we note that this structure
group fixes the four basis spinors $\e$, $J^7_{ij}\G^{ij}\e$,
$\G^7\e$ and $\G^8\e$. It is thus consistent with the existence of 2,
3 or 4 Killing spinors. 

\paragraph{$(G_2\ltimes\mbb^7)\times\mbb^2$ structure, N=2.}
Suppose that we demand, instead of the most generic (of the form
discussed in this subsection) second Killing spinor $\s_{(1)}$ of
the previous paragraph, that $f^A=0$ in (\ref{g2g2}). In this case, we
may take the second Killing spinor to be
\be
\a=(f_{(1)}+u_{(1)}\G^8)\e,
\end{equation}
and the existence of this Killing spinor implies that the structure
group is reduced to $(G_2\ltimes\mbb^7)\times\mbb^2$; the $G_{2}$ acts
nontrivially on the 1234567 directions. 
It is straightforward to
verify that in this case
$\a$ is annihilated by (\ref{lie}) if and only if
$f^{AB}K^{AB}_{i8}=0$. These are seven independent conditions, which
agrees with the fact that the Lie algebra of $G_2$ is fourteen
dimensional. This structure group stabilises two basis spinors, and is
thus consistent with $N=2$.

\paragraph{$\mbb^9$ structure, N=3,4,...,16.} 
Next let us suppose that in addition to the
generic (of the form discussed in this subsection) pair of Killing
spinors $\e$, $\s_{(1)}$, we demand the existence of a generic third
Killing spinor
\be
\s_{(2)}=\eta_{SU(3)\ltimes}+(f_{(2)B}\G^{B8}+u_{(2)B}\G^B)\e,
\end{equation}
where $\eta_{SU(3)\ltimes}$ is an arbitrary spinor fixed by the
$(SU(3)\ltimes\mbb^6)\times\mbb^3$ given above,
$f_{(2)B}$, $u_{(2)B}$, $B=1,...,6$  are arbitrary generic non-zero
one forms in the $\mathbf{3}+\overline{\mathbf{3}}$ of 
$SU(3)$ and we have reexpressed $J^A_{ij}\G^{ij}\e$ using
(\ref{annh}). We may exploit the fact that $SU(3)$ acts transitively
on the unit sphere in its $\mathbf{3}+\overline{\mathbf{3}}$, 
and that
\be
\frac{SU(3)}{SU(2)}\cong S^5
\end{equation}
to set 
\be\label{oop}
\s_{(2)}=\eta_{SU(3)\ltimes}+(f_{(2)}\G^{68}+u_{(2)6}\G^6+u_{(2)5}\G^{5}+u_{(2)D}\G^D)\e,
\end{equation}
where $D=1,...4$ and $u_{(2)D}$ is a non-zero one form in the
$\mathbf{2}+\overline{\mathbf{2}}$ of $SU(2)$. We still have the freedom to
perform $SU(2)$ transformations in the 1234 directions preserving
$f_{(2)B}=\d_{6B}f_{(2)}$, $B=1,...,6$. We may use the fact that $SU(2)$ acts
transitively on the unit sphere in its
$\mathbf{2}+\overline{\mathbf{2}}$, and 
\be
SU(2)\cong S^3,
\end{equation}
to write $\s_{(2)}$ as
\be
\s_{(2)}=\eta_{SU(3)\ltimes}+(f_{(2)}\G^{68}+u_{(2)6}\G^6+u_{(2)5}\G^{5}+u_{(2)4}\G^4)\e.
\end{equation}  
and deduce that the existence of the Killing spinors $\e$, $\s_{(1)}$, and $\s_{(2)}$ implies that the
structure group is reduced to $\mbb^9$. These Killing spinors are  
annihilated by (\ref{lie}) if and only if $f^{AB}K^{AB}_{ij}=0$. An $\mbb^9$ 
structure group fixes the sixteen basis spinors $\e$,
$J^A_{ij}\G^{ij}\e$, $\G^i\e$, and is thus consistent with $N=3,...,16$.

\paragraph{$(SU(2)\ltimes\mbb^4)\times\mbb^5)$ structure, N=3,4,...,8.}
Let us suppose that instead of the generic third Killing spinor
$\s_{(2)}$ (of the form discussed in this subsection) there exists a
third Killing spinor with $u_{(2)D}=0$, $D=1,...,4$ in (\ref{oop}). Then demanding that
this non-generic third Killing spinor is annihilated by (\ref{lie}) in
addition to the generic first two implies that in addition to the conditions
$f^{AB}K^{AB}_{i8}=f^{AB}K^{AB}_{i7}=0$, we must have the five further
conditions  
$f^{AB}K^{AB}_{i6}=0$. Note that these conditions identically imply that
$f^{AB}K^{AB}_{i5}=0$. The existence of a third Killing spinor of this
non-generic form thus implies that the structure group is reduced to
$(SU(2)\ltimes\mbb^4)\times\mbb^5$, where the $SU(2)$ acts
nontrivially in the 1234 directions. This structure group stabilises
the eight basis spinors $\e$, $J^H_{ij}\G^{ij}\e$, $\G^H\e$,
$H=5,6,7$, and is thus consistent with $N=3,...,8$.

\subsection{Additional Killing spinors of the form
  $(f+\frac{1}{8}f^AJ^A_{ij}\G^{ij}+g\G^-+\frac{1}{8}g^AJ^A_{ij}\G^{-ij})\e$}
In this subsection we will examine the reduction of the structure
group implied by incorporating progressively more Killing spinors of
the form 
\be\label{xxx}
(f+\frac{1}{8}f^AJ^A_{ij}\G^{ij}+g\G^-+\frac{1}{8}g^AJ^A_{ij}\G^{-ij})\e.
\end{equation}
The
structure groups which can arise (distinct from those of the
subsection before last, which are obtained  with $g=g^A=0$), and the number of
Killing spinors of this form they can fix, are summarised as follows:
\begin{center}
\begin{tabular}{|c|c|}\hline$G$&$N$\\\hline\hline
  Spin(7)&2\\\hline 
  $SU(4)$&2,3,4\\\hline $Sp(2)$&2,3,...,6\\\hline
  $SU(2)\times SU(2)$&3,4,...,8\\\hline Chiral
  $SU(2)$&3,4,...,10\\\hline $U(1)$&4,5,...,12\\\hline Chiral Identity&4,5...16\\\hline 
\end{tabular}
\end{center}
Let us consider each case in more detail. We reserve the notation
$\xi$ for Killing spinors which are completely generic among those
discussed in this subsection. $\xi_{(i)}$ denotes the $i$th such spinor.
Note that for any configuration admitting multiple Killing spinors of
the form (\ref{xxx}), with at least one of $g,g^A\neq0$, we have the
freedom to act with the $(Spin(7)\ltimes\mbb^8)\times\mbb$ element
\be
1+q\G^{+9},
\end{equation}
so that by an appropriate choice of $q$ one of the Killing spinors may
be chosen such that one of the $f,f^A=0$, since $\e$ satisfies
\be\label{jkl}
\G^{+-}\e=\G^9\e=\e.
\end{equation}  

\paragraph{$Spin(7)$ structure, N=2.}
Suppose we demand, in addition to $\e$, the existence of the non-generic second Killing spinor
\be
\a=(f+g\G^-)\e.
\end{equation}
The existence of this Killing spinor implies that the structure group
is reduced to Spin(7). $\a$ is annihilated by (\ref{lie}) if and only
if $\theta_i=q=0$.

\paragraph{$SU(4)$ structure, N=2,3,4.}
Instead of $\a$, suppose we demand the existence of the non-generic second
Killing spinor
\be
\b=\eta_{Spin(7)}+\frac{1}{8}f^AJ^A_{ij}(g\G^{ij}+h\G^{-ij})\e,
\end{equation}
where $\eta_{Spin(7)}$ is an arbitrary spinor fixed by the Spin(7) of
the previous paragraph, $f^A\neq0$ and $\G^+\b\neq0$. Since a Spin(7)
transformation on the spinor irreps commutes with $\G^-$, 
we can perform a Spin(7) transformation to set
\be
\b=\eta_{Spin(7)}+\frac{1}{8}fJ^7_{ij}(g\G^{ij}+h\G^{-ij})\e,
\end{equation}
and this is annihilated by (\ref{lie}) if and only if
$f^{7A}=\theta_i=q=0$. The stabiliser of $\e$ and $\b$ is thus $SU(4)$, and
since this group fixes the four basis spinors $\e$, $\G^-\e$,
$J^7_{ij}\G^{ij}\e$, $J^7_{ij}\G^{-ij}\e$, it is consistent with
$N=2,3,4$.

\paragraph{$Sp(2)$ structure, N=2,3,...,6.}
Instead of $\a$ or $\b$, suppose we demand the existence of the most
generic (of the form discussed in this subsection) second Killing
spinor $\xi_{(1)}$, 
\be
\xi_{(1)}= \eta_{Spin(7)}+(\frac{1}{8}f^AJ^A_{ij}\G^{ij}+\frac{1}{8}g^AJ^A_{ij}\G^{-ij})\e,
\end{equation}
where $f^A$, $g^A$ are non-zero and not linearly dependent. Then from
subsection 3.1, we may set
\be
\xi_{(1)}=\eta_{SU(4)}+\frac{1}{8}f_{(1)}J^6_{ij}\G^{ij}\e,
\end{equation}
where $\eta_{SU(4)}$ is an arbitrary spinor fixed by the $SU(4)$ of
the previous paragraph. It is easy to verify that the stabiliser of $\e$ and
$\xi_{(1)}$ is $Sp(2)$. Since this group stabilises the six basis
spinors $\e$, $\G^-\e$,
$J^7_{ij}\G^{ij}\e$, $J^7_{ij}\G^{-ij}\e$, $J^6_{ij}\G^{ij}\e$, $J^6_{ij}\G^{-ij}\e$, it is consistent
with $N=2,3,...,6$. 

\paragraph{$SU(2)\times SU(2)$ structure, N=3,4,...,8.}
In addition to the two most generic (of the form discussed in this
subsection) Killing spinors $\e$, $\xi_{(1)}$, suppose we demand the
existence of the non-generic third Killing spinor
\be
\gamma=\eta_{Sp(2)}+\frac{1}{8}f^CJ^C_{ij}(g\G^{ij}+h\G^{-ij})\e,
\end{equation}
where $\eta_{Sp(2)}$ is an arbitrary spinor fixed by the $Sp(2)$ of
the previous paragraph, $C=1,...,5$, $f^C\neq0$ and
$\G^+\gamma\neq0$. We may set
\be
\gamma=\eta_{Sp(2)}+\frac{1}{8}fJ^5_{ij}(g\G^{ij}+h\G^{-ij})\e,
\end{equation}
and deduce that the existence of this third Killing spinor $\gamma$
implies that the structure group is reduced to $SU(2)\times SU(2)$. This group fixes eight
of the basis spinors, so is consistent with $N=3,...,8$.

\paragraph{Chiral $SU(2)$, N=3,4,...,10.}
Intead of $\gamma$, suppose we demand, in addition to $\e$ and $\xi_{(1)}$, the
existence of the most generic (of the form of this subsection) third
Killing spinor $\xi_{(2)}$. We may take this to be
\be
\xi_{(2)}=\eta_{SU(2)\times SU(2)}+\frac{1}{8}f_{(2)}J^4_{ij}\G^{ij}\e,
\end{equation}
where $\eta_{SU(2)\times SU(2)}$ is an arbitrary spinor fixed by the
$SU(2)\times SU(2)$ of the previous paragraph. In addition to the
other two, this Killing spinor is fixed by
an $SU(2)$ acting non-trivially in eight dimensions. We refer to this
structure as a chiral $SU(2)$ structure, and the same remarks made for
the case of the $(SU(2)\ltimes\mbb^8)\times\mbb$ structure apply in
this case. Since this $SU(2)$ fixes ten basis spinors, it is
consistent with $N=3,...,10$.

\paragraph{$U(1)$ structure, N=4,...,12}
Suppose we demand, in addition to the Killing spinors $\e$, $\xi_{(1)}$
and $\xi_{(2)}$, the fourth non-generic Killing spinor
\be
\d=\eta_{CSU(2)}+\frac{1}{8}f^EJ^E_{ij}(g\G^{ij}+h\G^{-ij})\e,
\end{equation}
where $\eta_{CSU(2)}$ is an arbitrary spinor fixed by the chiral $SU(2)$ of
the previous paragraph, $E=1,2,3$, $f^E\neq0$ and 
$\G^+\d\neq0$. We may set
\be
\d=\eta_{CSU(2)}+\frac{1}{8}fJ^3_{ij}(g\G^{ij}+h\G^{-ij})\e,
\end{equation}
and in addition to the other three Killing spinors, this is fixed by $U(1)$. Since this group fixes twelve basis
spinors, it is consistent with $N=4,...,12$. The same remarks made for
$(SU(2)\ltimes\mbb^8)\times\mbb$ apply in this case.

\paragraph{Chiral Identity structure, N=4,...,16}
Instead of $\d$, suppose we demand, in addition to $\e$, $\xi_{(1)}$
and $\xi_{(2)}$, the existence of the most generic (of the form discussed in this
subsection) fourth Killing spinor. We may take this to be of the form 
\be
\xi_{(3)}=\eta_{U(1)}+\frac{1}{8}f_{(3)}J^2_{ij}\G^{ij}\e,
\end{equation}
where $\eta_{U(1)}$ is an arbitrary spinor fixed by the $U(1)$ of the
previous paragraph. The existence of $\xi_{(3)}$ in addition to the
other three Klling spinors reduces the
structure group to the identity. An identity structure is thus
consistent with 4,...,16 Killing spinors of the same chirality on the
base. 

\subsection{Additional Killing spinors of generic form}
Finally let us discuss the case where the additional Killing spinors
are assumed to be of the most generic form,
\be\label{yyy}
(f+\frac{1}{8}f^AJ^A_{ij}\G^{ij}+u_i\G^i+g\G^-+\frac{1}{8}g^AJ^A_{ij}\G^{-ij}+v_i\G^{-i})\e.
\end{equation}
The new structure groups we find in this case, together with the
number of Killing spinors they can fix, are as follows:  
\begin{center}
\begin{tabular}{|c|c|}\hline$G$&$N$\\\hline\hline
  $G_2$&2,3,4\\\hline 
  $SU(3)$&2,3,...,8\\\hline $SU(2)$&2,3,...,16\\\hline Identity&3,4,...32\\\hline 
\end{tabular}
\end{center} 
 
\paragraph{$SU(2)$ structure, N=2,3,...,32.}
Before we obtain the structure group defined by the most generic
second Killing spinor, consider first a second Killing spinor of the
form (\ref{yyy}), but with $v_i=0$, and at least one of
$g,g^A\neq0$. We may act on this Killing spinor with the
$(Spin(7)\ltimes\mbb^8)\times\mbb)$ element
\be
1+\theta_i\G^{+i},
\end{equation}
where
\be
\theta_{i}=\frac{1}{(g^2+g^Ag^A)}(gu_i+g^AJ^A_{ij}u^j).
\end{equation}
Given the projections (\ref{jkl}) satisfied by $\e$, together with
\bea
\G_{ijk}\e&=&-\phi_{ijkl}\G^l\e,\nn
\phi_{ijkl}J^{Akl}&=&-6J^A_{ij},
\eea
we see that we may set $u_i=0$, and thus take the Killing spinor to be
of the form discussed in the previous subsection. So now consider the case of the most generic second Killing
spinor, given by (\ref{yyy}) with all thirty-two functions (and in
particular, the $v_i$) non-zero. We can act on the $\mathbf{8}_-$ spanned by the $\G^{-i}\e$ with Spin(7)
to set these components to be of the form
\be\label{mkan}
v\G^{-8}\e.
\end{equation}
We have the freedom to perform $(G_2\ltimes\mbb^7)\times\mbb^2$ transformations preserving this
form. By acting with
\be
1+\theta_A\G^{+A}+\theta_8\G^{+8}+g\G^{+9},
\end{equation}
we may set $f=f^A=u_8=0$, and we still have the freedom to perform
$G_2$ transformations. We exploit this to set the $\mathbf{7}_{G_2}$
spanned by the $\G^A\e$ to be
\be\label{mka}
u\G^7\e,
\end{equation}
and we still have the freedom to perform $SU(3)$ transformations
preserving (\ref{mkan}), (\ref{mka}), and $f=f^A=0$. We exploit this to set the
components in the $\mathbf{7}=\mathbf{1}+(\mathbf{3}+\bar{\mathbf{3}})$ spanned by the
$J^A_{ij}\G^{-ij}$ to be of the form
\be
\frac{1}{8}g^GJ^G_{ij}\G^{-ij}\e,
\end{equation} 
where $G=6,7$. We have now completely fixed the most generic second
Killing spinor; it takes the form 
\be\label{generic}
\lambda_{(1)}=(u\G^7+g\G^-+\frac{1}{8}g^GJ^G_{ij}\G^{-ij}+v\G^{-8})\e,
\end{equation}
where $G=6,7$. Thus, the common stabiliser of the most
generic $\lambda_{(1)}$ and $\e$ is $SU(2)$, acting on the 1234
directions. Such a group fixes sixteen basis spinors, and so is
consistent with $N=2,3,...,16$.

\paragraph{$SU(3)$ structure, N=2,3,...,8.}
Suppose that in (\ref{generic}) $g^6=0$. Then
$\lambda_{(1)}$ and $\e$ are
stabilised by an $SU(3)$ acting in the 123456
directions. This structure group stabilises eight basis spinors, and
is thus consistent with $N=2,3,...,8$.

\paragraph{$G_2$ structure, N=2,3,4.}
Suppose that in (\ref{generic}), $g^6=g^7=u=0$. Then $\lambda_{(1)}$ and $\e$ are
stabilised by a $G_2$ acting in the 1234567
directions. This structure group stabilises four basis spinors, and
is thus consistent with $N=2,3,4$. 

\paragraph{Identity structure, $N=3,4,...,32$}
Incorporating a third generic Killing spinor in addition to $\e$ and the
generic $\lambda{(1)}$ reduces the structure group to the identity. An
identity structure is thus consistent with $N=3,4,...,32$. This
completes the classification of subgroups of
$(Spin(7)\ltimes\mbb^8)\times\mbb$ which can arise as structure groups
in eleven dimensions.

\section{Structures which embed both in
  $(Spin(7)\ltimes\mbb^8)\times\mbb$ and in $SU(5)$}
In the previous section we classified all structure groups which can
arise as subgroups of $(Spin(7)\ltimes\mbb^8)\times\mbb$ - that is,
all groups which can arise as structure groups of eleven dimensional
spacetimes admitting at least one null Killing spinor. When the
analysis begun in \cite{pap} is completed, 
the analagous question will be answered for all eleven dimensional
spacetimes admitting at least one timelike Killing spinor, where the
structure group is a subgroup of $SU(5)$. There can clearly be a
redundancy in the classification for structure groups that are
subgroups both of $(Spin(7)\ltimes\mbb^8)\times\mbb$ and of $SU(5)$;
some (but not all) spacetimes with these structure groups can admit both
timelike and null Killing spinors. To lift this redundancy, we will
make the following definitions:
\\\\
A supersymmetric spacetime is said to admit a timelike G-structure if
it admits only timelike Killing spinors.
\\\\
A supersymmetric spacetime is said to admit a mixed G-structure if it
admits both timelike and null Killing spinors.
\\\\
A supersymmetric spacetime is said to admit a null G-structure if it
admits only null Killing spinors.  
\\\\
Spacetimes admitting timelike G-structures may be classified  using the framework of
\cite{gaunt1}, \cite{pap}, and spacetimes admitting null
G-structures can be classified
using the framework of \cite{gaunt3} and this paper. Spacetimes
admitting mixed
G-structures can be classified using either. We now want to determine
which of the G-structures of the previous section are null and which
are mixed. Those
G-structures for which G is not a subgroup of $SU(5)$ are trivially
null. A complete list of such groups is given by
$(Spin(7)\ltimes\mbb^8)\times\mbb$ itself, all the groups of
subsection 3.1, all the groups of subsection 3.2, Spin(7), and
$G_2$. The remaining groups of subsections 3.3 and 3.4 are subgroups
of both $(Spin(7)\ltimes\mbb^8)\times\mbb$ and $SU(5)$, so we must
analyse these cases in more detail. 

Let us thus assume that $\e$ and $\chi$ are Killing, where
$\chi$ is assumed to be of the general form
\be\label{gennn}
\chi=(f+\frac{1}{8}f^AJ^A_{ij}\G^{ij}+u_i\G^i+g\G^-+\frac{1}{8}g^AJ^A_{ij}\G^{-ij}+v_i\G^{-i})\e.  
\end{equation}
The fiducial Killing spinor $\e$ is null. If $\chi$ is null, it
may still be that the linear combination $\chi+\rho\e$ is timelike,
for some $\rho\in\mbb$. Thus the G-structure defined by $\e$ and
$\chi$ is null if and only if the Killing spinor
$\chi+\rho\e$ is null for all $\rho\in\mbb$. Let us compute the
one form $V$ with components
\be
V_{\m}=(\overline{\chi+\rho\e})\G_{\m}(\chi+\rho\e),
\end{equation}
by setting $f\rightarrow f+\rho$ in (\ref{gennn}). We find
\bea
V&=&((f+\rho)^2+u_iu^i+f^Af^A)e^+-2(g^2+v_iv^i+g^Ag^A)e^-\nn&+&2(v_iu^i-(f+\rho)g-f^Ag^A)e^9
+2(g^Au^iJ^A_{ij}+f^Av^iJ^A_{ij}-(f+\rho)v_j-gu_j)e^j.
\eea
Now, using the identity
\be
u^iJ^A_{ij}v^ju^kJ^A_{kl}v^l=u_iu^iv_jv^j-(u_iv^i)^2,
\end{equation}
we find that
\be
V_{\m}V^{\m}=-[(f+\rho)g^A-gf^A+u^iJ^A_{ij}v^j][(f+\rho)g^A-gf^A+u^kJ^A_{kl}v^l]-2f^{[A}g^{B]}f^Ag^B.
\end{equation}
Thus, $V_{\m}V^{\m}=0$ for all $\rho\in\mbb$ if and only if
\bea
\label{bag}g^A&=&0,\\\label{bagg}
gf^A&=&u^iJ^A_{ij}v^j,
\eea
and these are the necessary and sufficient conditions on the spinor
$\chi$ for the G-structure defined by $\e$, $\chi$ to be
null. A G-structure which is defined by more than two Killing
spinors is null if and only if these conditions hold for all linear
combinations of the Killing spinors with constant coefficients. 

The G-structures with structure groups $SU(4)$, $Sp(2)$,
$SU(2)\times SU(2)$, Chiral $SU(2)$, $U(1)$ or Chiral Identity are
never null, since they always admit Killing spinors with
$g^A\neq0$, or $gf^A\neq0$, $u^i=v^i=0$. Thus, if they embed in
$(Spin(7)\ltimes\mbb^8)\times\mbb$, these structure groups may always
be embedded in $SU(5)$ as well, and these G-structures are thus mixed.
It will be most efficient to
analyse these cases using the timelike basis of spinors of
\cite{pap}, since one can also have timelike G-structures for these
groups. 

Finally we note that although $SU(3)$, $SU(2)$ and the Identity are
subgroups of both $(Spin(7)\ltimes\mbb^8)\times\mbb$ and $SU(5)$, it
is possible to have null $SU(3)$, $SU(2)$ or Identity structures if
(\ref{bag}), (\ref{bagg}) hold for all linear combinations of the
associated Killing spinors with constant coefficients. One may also
have mixed or timelike G-structures with these structure groups.

In summary, the G-structures which are always null in eleven
dimensions have structure groups
 
\begin{center}
\begin{tabular}{|c|}\hline$G$\\\hline\hline
  $(Spin(7)\ltimes\mbb^8)\times\mbb$\\\hline$(SU(4)\ltimes\mbb^8)\times\mbb$\\\hline $(Sp(2)\ltimes\mbb^8)\times\mbb$\\\hline
  $(SU(2)\times SU(2))\ltimes\mbb^8)\times\mbb$\\\hline 
  $(SU(2)\ltimes\mbb^8)\times\mbb$\\\hline $(U(1)\ltimes\mbb^8)\times\mbb$\\\hline
  $\mbox{Chiral}\;\;\mbb^9$\\\hline $(G_2\ltimes\mbb^7)\times\mbb^2$\\\hline 
  $(SU(3)\ltimes\mbb^6)\times\mbb^3$\\\hline $(SU(2)\ltimes\mbb^4)\times\mbb^5$\\\hline
  $\mbb^9$\\\hline$Spin(7)$\\\hline$G_2$\\\hline
\end{tabular}
\end{center} 
The G-structures which, although their structure groups may be embedded in
$(Spin(7)\ltimes\mbb^8)\times\mbb$, are always mixed or timelike, are
as follows:
\begin{center}
\begin{tabular}{|c|}\hline$G$\\\hline\hline
  
  $SU(4)$\\\hline $Sp(2)$\\\hline
  $SU(2)\times SU(2)$\\\hline Chiral
  $SU(2)$\\\hline $U(1)$\\\hline Chiral Identity\\\hline 
\end{tabular}
\end{center}
The G-structures which can be null if and only if (\ref{bag}), (\ref{bagg})
hold for all linear combinations of the associated Killing spinors
with constant coefficients, and which can otherwise be mixed or
timelike, have structure groups 
\begin{center}
\begin{tabular}{|c|}\hline$G$\\\hline\hline 
  $SU(3)$\\\hline $SU(2)$\\\hline Identity\\\hline 
\end{tabular}
\end{center} 

\section{Conclusions}
In this work we have classified all structure groups which can arise
as subgroups of $(Spin(7)\ltimes\mbb^8)\times\mbb$ in eleven
dimensions, and we have constructed the spaces of spinors they
fix. We have introduced the notions of timelike, mixed and null
G-structures for 
supersymmetric spacetimes. We
have further classified the structure subgroups of
$(Spin(7)\ltimes\mbb^8)\times\mbb$ according to these definitions. 

The objective of the refined G-structure classification scheme in
this context is to categorize the general local bosonic solution of the
Killing spinor equation of eleven dimensional supergravity. We believe
that this is a viable objective. The idea is ultimately to cover all
possible multi-spinor ans\"{a}tze for the Killing spinor equation, in
full generality. In
this work, the
different types of 
  ansatz assuming at least one null Killing spinor have been
  classified, with the 
  common isotropy group of the spinors providing the organising
  principle. Since the spinors are assumed to be Killing, their common
  isotropy group is elevated to the status of the
  structure group of the supersymmetric spacetimes admitting those
  Killing spinors. 

Now that the possible types of multi-spinor ans\"{a}tze with at least
one null spinor have been
classified, it remains to perform a direct analysis of the Killing
spinor equation. As discussed in the introduction, this will proceed by
imposing
\be\label{opo}
[\mathcal{D}_{\m},Q]\e=0,
\end{equation}
for the most general choice of $N$ distinct $Q$s for every $N$ for
every structure group discussed in this paper and every structure group arising
in the timelike case (though in certain cases, we expect that the
analysis of the Killing spinor equation will reveal that the
existence of a given set of Killing spinors will identically imply the
existence of more, thus reducing the number of distinct cases to be
covered). Nevertheless, to see this project through to completion will clearly
involve a massive computational effort. 

In \cite{spin7}, we have begun with the case of a Spin(7) structure,
and have derived the constraints implied by imposing
\be
[\md_{\m},g\G^-]\e=0.
\end{equation}
We have found that the the general bosonic solution of the Killing spinor
equation admitting a Spin(7) structure is determined locally as
follows.  We may
take the Killing spinors to be $\e$, $H^{-1/3}(x)\G^-\e$, with
metric
\bea
ds^2&=&H^{-2/3}(x)\Big(2[du+\lambda(x)_Mdx^M][dv+\v(x)_Ndx^N]+[dz+\s(x)_Mdx^M]^2\Big)\nn&+&H^{1/3}(x)h_{MN}(x)dx^Mdx^N,
\eea
where $h_{MN}$ is a metric of Spin(7) holonomy and $d\lambda$, $d\v$
and $d\s$ are two-forms in the $\mathbf{21}$ of Spin(7). Observe that
there are three Killing vectors. Defining the
elfbeins 
\bea
e^+&=&H^{-2/3}(du+\lambda),\nn e^-&=&dv+\v,\nn
e^9&=&H^{-1/3}(dz+\s),\nn e^i&=&H^{1/6}\hat{e}^i(x)_Mdx^M,
\eea
where $\hat{e}^i$ are the achtbeins for $h$, the four-form is
\bea
F&=&e^+\wedge e^-\wedge e^9\wedge d\log H+H^{-1/3}e^+\wedge e^-\wedge
d\s-e^+\wedge e^9\wedge d\v\nn&+&H^{-2/3}e^-\wedge e^9\wedge
d\lambda+\frac{1}{4!}F^{\mathbf{27}}_{ijkl}\hat{e}^i\wedge\hat{e}^j\wedge\hat{e}^k\wedge\hat{e}^l.
\eea

We expect that classifying all supersymmetric spacetimes with
G-structures whose structure groups fix at most eight Killing spinors
will be eminently practical. Considerably more effort will be required to
completely classify the cases ($\mbb^9$ and $SU(2)$) where the
structure group fixes at most sixteen. Finally, the classification of
all spacetimes admitting an Identity structure will be, without further
insight, technically difficult. This is because the Killing
spinors in this case can be completely generic, and the constraints
implied by the existence of a completely generic Killing spinor will be very complicated. It would be very useful to
have a complementary ``top down'' formalism, whereby one could start
with the constraints for maximal supersymmetry, and progressively and
systematically weaken them, for the classification of spacetimes
admitting an Identity structure and preserving $N>16$. Nevertheless,
we believe that classifying all spacetimes with an Identity structure
will be a hard problem, irrespective of the approach employed. 

Clearly, the formalism we have used is universally applicable. Simpler
lower dimensional supergravites (particularly those with eight
supercharges) should be easy to analyse. It should also be possible to
apply it to IIB, though given the formal complexity of this theory,
the amount of computation required to perform the complete
classification will, if anything, be greater than that in eleven
dimensions (though there is the simplification that in IIB the spinors are
always null \cite{HJ}, and all structure groups are subgroups of
$Spin(7)\ltimes\mbb^8$). 

Of course, categorizing the general local bosonic solution of the Killing spinor
equation of eleven dimensional supergravity is not the same as
determining all the supersymmetric solutions of the field equations
and Bianchi identity; some subset thereof will still need to be
imposed. Nevertheless, it is our hope that the procedure we are
advocating will ultimately lead to the construction of many
new solutions, from which much will be learned about the physics of
supersymmetric spacetimes in M-theory.

\section{Acknowledgements}
We are grateful to Jerome Gauntlett and Gary Gibbons for useful
comments.  M. C. is
supported by 
EPSRC, Cambridge European Trust and Fondazione Angelo Della Riccia. OC
is supported by a Senior Rouse Ball Scholarship.

\end{document}